\documentclass[twocolumn,english,notitlepage,superscriptaddress]{revtex4-1}
\usepackage[T1]{fontenc}
\usepackage[latin9]{inputenc}
\usepackage{amsmath}
\usepackage{amssymb}
\usepackage{graphicx}

\usepackage{bibunits}
\defaultbibliography{ch.bib}
\defaultbibliographystyle{unsrt}

\makeatletter
\usepackage{babel}

\makeatother

\usepackage{babel}
\begin{document}
\begin{bibunit}
\title{Revealing and concealing entanglement with non-inertial motion}

\author{Marko Toro\v{s}}
\email{m.toros@ucl.ac.uk}

\selectlanguage{english}%

\affiliation{Department of Physics and Astronomy, University College London, Gower
Street, WC1E 6BT London, UK}

\author{Sara Restuccia}

\affiliation{School of Physics and Astronomy, University of Glasgow, Glasgow,
G12 8QQ, UK}

\author{Graham M. Gibson}

\affiliation{School of Physics and Astronomy, University of Glasgow, Glasgow,
G12 8QQ, UK}

\author{Marion Cromb}

\affiliation{School of Physics and Astronomy, University of Glasgow, Glasgow,
G12 8QQ, UK}

\author{Hendrik Ulbricht}

\affiliation{Department of Physics and Astronomy, University of Southampton, SO17
1BJ, UK}

\author{Miles Padgett}

\selectlanguage{english}%

\affiliation{School of Physics and Astronomy, University of Glasgow, Glasgow,
G12 8QQ, UK}

\author{Daniele Faccio}
\email{daniele.faccio@glasgow.ac.uk}

\selectlanguage{english}%

\affiliation{School of Physics and Astronomy, University of Glasgow, Glasgow,
G12 8QQ, UK}
\begin{abstract}
Photon interference and bunching are widely studied quantum effects
that have also been proposed for high precision measurements. Here we construct a theoretical description of
photon-interferometry on rotating platforms, specifically exploring the relation between non-inertial
motion, relativity, and quantum mechanics. On the basis of this, we then propose an experiment where photon entanglement can be revealed or concealed solely by controlling
the rotational motion of an interferometer, thus providing  a
 route towards studies at the boundary between quantum mechanics and relativity. 

\end{abstract}
\maketitle

{\bf{Introduction}}.
The notions of space and time are at the core of modern physics and
remain an area of intense research~\cite{minkowski1909raum,petkov2010minkowski}.
A striking example of how elementary notions of space and time lead to
surprising consequences is the derivation of Lorentz transformations,
a cornerstone of quantum field theory, utilizing only basic assumptions~\cite{von1910einige,liberati2013tests}.


The exploration of the special-relativistic regime is historically
strongly linked to investigations of the propagation of light~\cite{robertson1949postulate},
e.g. the Michelson\textendash Morley experiment~\cite{michelson1887relative}.
More recent experiments have also started to probe the quantum nature of
light, e.g. the Hong-Ou-Mandel (HOM) experiment~\cite{hong1987measurement},
indirectly testing the underpinning spacetime symmetries. Quantum
optical interference effects, either one-photon or two-photon, are
thus of fundamental importance~\cite{mandel1999quantum}, as well
as providing paths to technological applications~\cite{lyons2018attosecond}.

A further test of special relativity, moving towards the domain and
ideas of general relativity, is possible in situations where linear
acceleration or rotational motion is present~\cite{gourgoulhon2016special}.
A notable example is the classical Sagnac experiment where an interferometer
is placed on a rotating platform~\cite{sagnac1913ether,sagnac1913preuve,post1967sagnac}.
More recent experiments include experimental tests of photonic entanglement
in accelerated reference frames~\cite{fink2017experimental}, the demonstration of how to overcome the shot-noise limit
using an entanglement-enhanced optical gyroscope~\cite{fink2019entanglement} and the extension of HOM interference to rotating platforms~\cite{restuccia2019photon}. 

In this letter, we propose a new experimental platform based on the Mach-Zehnder interferometer that
explores the relation between interference, entanglement, and non-inertial
rotational motion. In particular, we discover that by simply setting
the apparatus in rotational motion one can detect or conceal entanglement.
We first provide a theoretical description of quantum experiments
on a rotating platform 
starting from the Hamiltonian on a generic Hilbert
space which we then apply to study photon-interferometry experiments.
The model also recovers the results for the Sagnac effect in the quantum regime~\cite{bertocchi2006single}
as well as for the recent demonstration of photon bunching in a rotating
reference frame~\cite{restuccia2019photon}. 

\begin{figure}
\includegraphics[width=1\columnwidth]{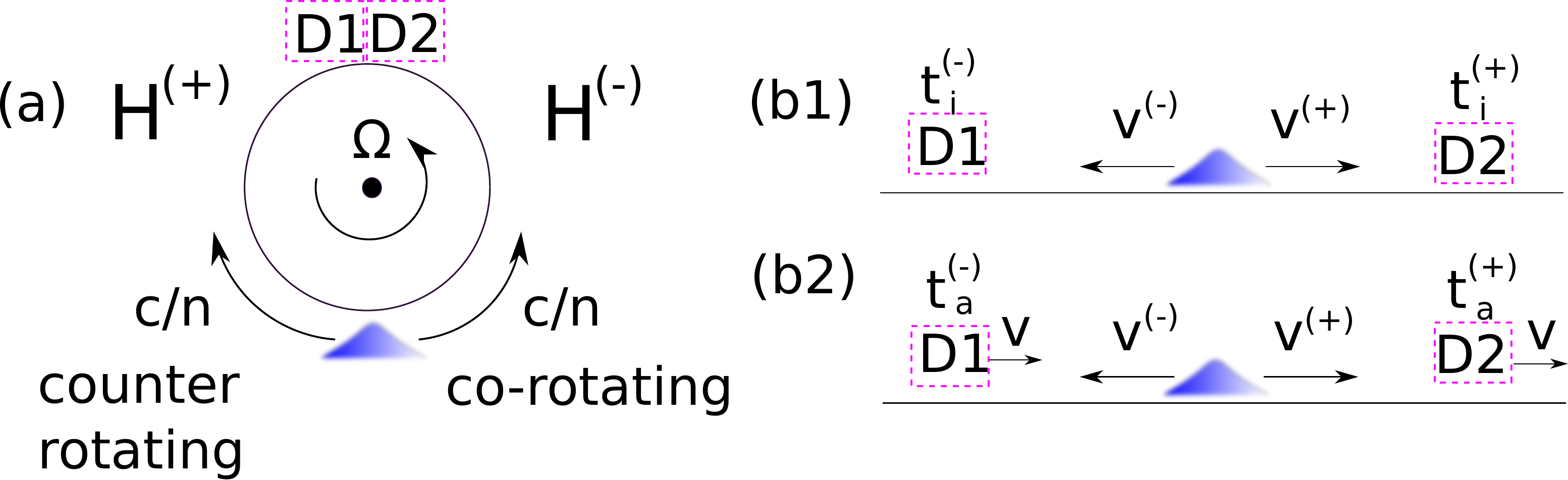} \caption{Conceptual setup. (a) Description from the viewpoint of the co-rotating
observer. The counter-rotating (co-rotating) quantities are shown
on the left (right). The counter-rotating (co-rotating) Hamiltonians
are different, while the light-speed in the two directions is equal (${c}/{n}$). D1 (D2) indicate the detectors for
the counter-rotating (co-rotating) direction, respectively. (b) Description from the inertial
laboratory frame represented in a straight line (see supplemental
material). Here both the Hamiltonian and the speed of light differ
in the two directions, i.e. $v^{(-)}\protect\neq v^{(+)}$ and $v^{(\pm)}\protect\neq\frac{c}{n}$.
(b1) Only the light-drag effect is taken into account; $t_{i}^{(+)}$,
$t_{i}^{(-)}$ denote how long it would take to reach the detectors
assuming they would not have moved; the subscript $i$ stands for
``initial''. (b2) Both light-drag effect and the motion of detectors
is taken into account. $t_{a}^{(+)}$, $t_{a}^{(-)}$ denote how long
it takes to reach the detectors; the subscript $a$ stands for ``actual''.
$v=r\Omega$ is the speed of the detectors as seen from the inertial
laboratory reference frame. }
\label{setup} 
\end{figure}

{\bf{Theoretical model}}.
We consider an experimental platform rotating at angular frequency
$\Omega$ depicted in Fig.~\ref{setup}(a). The system is confined
to move on a circle of radius $r$ in the equatorial plane normal to the rotation axis.
We further suppose there is a co-rotating medium with refractive
index $n$. For the co-rotating observer, the light propagation speed
is ${c}/{n}$ in both directions as can be deduced from symmetry considerations.
It is also instructive to describe the same experiment from the inertial
frame of the laboratory: one can formally map the circular trajectories
to straight-line motions as shown in Fig.~\ref{setup}(b)~\cite{tartaglia2015sagnac}.
In this latter case one has to account for the light-dragging effect
~\cite{post1967sagnac}, i.e. the Fizeau effect (see supplementary
material). 

To describe quantum mechanically the evolution of the system we choose
for convenience the viewpoint of the co-rotating observer (see Fig.~\ref{setup}(a)).
To account for the non-inertial motion
when $\Omega\neq0$ we start from the Born chart and exploit the methods
of symplectic Hamiltonian mechanics~\cite{da2001lectures,woit2017quantum}.
In particular, exploiting the so-called co-moment map from the generators
of the Poincaré algebra to Hilbert space operators one finds the following
Hamiltonian (see supplementary material): 
\begin{equation}
\hat{H}_{\text{Born}}=\Gamma(\hat{H}+r\Omega\hat{P}),\label{eq:bornRel}
\end{equation}
where $\Gamma=(1-(\frac{\Omega r}{c})^{2})^{-\frac{1}{2}}$ is the
Lorentz factor. The term $\sim\hat{P}$ keeps track of the non-inertial
motion of the detector, $\hat{P}$, which is the generator of translations,
changes the relative distance between the detector and the system,
e.g. in a time $\delta t$ the relative distance changes by $r\Omega\delta t$.

We now apply the Hamiltonian in Eq.~(\ref{eq:bornRel}) to photon-interferometry.
We use the Abraham relation between kinetic momentum and
energy~\cite{padgett2008diffraction,barnett2010resolution}: 
\begin{alignat}{1}
\hat{H}=nc\vert\hat{P}\vert,\label{eq:hp}
\end{alignat}
We combine Eqs.~(\ref{eq:bornRel}) and (\ref{eq:hp}) to find: 
\begin{equation}
\hat{H}_{\text{Born}}^{(\pm)}=\Gamma\left(1\pm\beta\right)\hat{H},\label{eq:bornRelA}
\end{equation}
where the positive (negative) superscript denotes the counter-rotating
(co-rotating) motion to be measured by the detector D1 (D2), and $\beta=\frac{r\Omega}{nc}$.
We note that Eq.~(\ref{eq:bornRelA}) suggests a simple physical interpretation
of the Hamiltonian as the \emph{relativistic Doppler-shifted energy}
of the system.

We can finally write the total Hamiltonian of the system: 
\begin{equation}
\hat{H}_{\text{Born}}=\hat{H}_{\text{Born}}^{(+)}\otimes\mathbb{I}+\mathbb{I}\otimes\hat{H}_{\text{Born}}^{(-)},\label{eq:h2}
\end{equation}
where we have assumed that the photons moving in opposite directions
do not interact, and $\mathbb{I}$ denotes the identity operators.
In summary, the total Hilbert space can be written as $\mathcal{H}=\mathcal{H}^{(+)}\otimes\mathcal{H}^{(-)},$
where $\mathcal{H}^{(+)}$($\mathcal{H}^{(-)}$) denotes the Hilbert
space of the counter-rotating (co-rotating) modes.

The time parameter $t$ that keeps track of the dynamics through Schrödinger's
equation is ticking as a clock following the detectors' motion; this
is a direct consequence of the quantization procedure that leads to
the Hamiltonian in Eq.~(\ref{eq:h2}). However, here we are mainly interested in the dominant effects where
one can approximate the Lorentz factor as $\Gamma\sim1$ and the
distances and times coincide to those measured by a ruler and a clock
in the inertial laboratory frame.

{\bf{Photon-interferometry experiments}}.
We now further develop the model by adopting Glauber's theory of
photo-detection~\cite{glauber1963quantum}. Here we will focus on
the experimental situation of photons with a coherence time that is short compared to the time resolution
of the detectors, but temporal aspects could be easily taken into
account~\cite{legero2006characterization,specht2009phase}.

To analyze photon-interferometry experiments we will work in the Schrödinger
picture~\cite{wang2006quantum}, where we will denote the initial
(final) state with the subscripts $i$ ($f$). We consider the experimental
situation where at time $t_{i}=0$ the photon is prepared in a state
$\vert\psi_{i}\rangle$, and then constrained to move in a circular
motion for a time $t_{f}=\frac{L}{c}n$ resulting in a state $\vert\psi_{f}\rangle$,
where $L$ is the traveled distance. Although one can always postulate
a given initial state it is nonetheless instructive to compare the
state $\vert\psi_{i}\rangle$, which is assumed to be generated by
the apparatus co-rotating with the platform, with the state generated
by the same apparatus when the platform is not rotating, i.e. when
$\Omega=0$. In particular, it is reasonable to assume that the frequencies
of the initial states generated in the two experimental situations
differ by at most $\sim\frac{\Omega}{2\pi}$. However, such a difference
\emph{in the initial state} produces only sub-leading effects which
are not amplified during \emph{time-evolution}, as can be explicitly
verified using the formulae we will develop. One can thus approximate
the initial states generated on the rotating platform with the states that would be generated at $\Omega=0$. 

The time-evolution is given by the usual Schrödinger equation with
the Hamiltonian in Eq.~\eqref{eq:h2}, i.e. 
\begin{alignat}{1}
\hat{H}^{(2)}= & \hbar\int d\omega\left[\omega^{(+)}\hat{a}^{\dagger}(\omega)\hat{a}(\omega)+\omega^{(-)}\hat{b}^{\dagger}(\omega)\hat{b}(\omega)\right],\label{eq:H2}
\end{alignat}
where we have defined $\omega^{(\pm)}=(1\pm\beta)\omega$, and $\hat{a}$
($\hat{b}$) is the counter (co-rotating) mode.

The state $\vert\psi_{f}\rangle$ then interferes at a beam-splitter
and one measures the outputs using two-detectors: the input modes
are $\hat{a}$ and $\hat{b}$ and we denote the output modes by $\hat{c}$
and $\hat{d}$. Here we consider the following relation between the
input and output modes: 
\begin{equation}
\left[\begin{array}{c}
\hat{c}(\omega)\\
\hat{d}(\omega)
\end{array}\right]=\frac{1}{\sqrt{2}}\left[\begin{array}{cc}
1 & 1\\
1 & -1
\end{array}\right]\left[\begin{array}{c}
\hat{a}(\omega)\\
\hat{b}(\omega)
\end{array}\right].\label{eq:bs}
\end{equation}
In particular, we are interested in the probability of detecting photons
in the modes $\hat{c}$ or $\hat{d}$. To this end, it is convenient
to define the temporal modes~\cite{blow1990continuum}: 
\begin{equation}
\hat{c}(t)=\mathcal{F}_{t}[\hat{c}(\omega)],\qquad\hat{d}(t)=\mathcal{F}_{t}[\hat{d}(\omega)],\label{eq:temporal}
\end{equation}
where $\mathcal{F}_{t}[\,\cdot\,]=\frac{1}{\sqrt{2\pi}}\int d\omega\,\cdot\,e^{-i\omega t}$.
In particular, we define the single-photon probability of detection
as 
\begin{equation}
P_{c}^{(1)}=\int dt\langle\psi_{f}\vert\hat{c}^{\dagger}(t)\hat{c}(t)\vert\psi_{f}\rangle,\label{eq:P1}
\end{equation}
with a similar definition for the probability $P_{d}^{(1)}$ for the
output mode $\hat{d}$. In addition, we also define the the two-photon probability of detection 
\begin{equation}
P^{(2)}=\int dt_{1}\int dt_{2}\langle\psi_{f}\vert \hat{d}^{\dagger}(t_{1})\hat{c}^{\dagger}(t_{2})\hat{c}(t_{2})\hat{d}(t_{1})\vert\psi_{f}\rangle,\label{eq:P2}
\end{equation}
which gives the coincidence probability. For the case $P^{(2)}<0.5$
 we speak of coalescence or HOM photon bunching and for $P^{(2)}>0.5$ we speak of photon anti-coalescence or anti-bunching. Classically,
one is limited to values $0.25<P^{(2)}<0.5$, making coincidence probabilities
a valuable tool to assess the quantum nature of the electromagnetic
field.  Importantly, anti-symmetrization and photon anti-coalescence reveals
hidden entanglement, as has already been demonstrated experimentally
in a non-rotating setup~\cite{fedrizzi2009anti}.

We consider first the experimental situation with a generic single-photon
input state: 
\begin{equation}
\vert\psi_{f}\rangle=\int d\omega\,\left[\psi_{a,f}(\omega)\hat{a}^{\dagger}(\omega)+\psi_{b,f}(\omega)\hat{b}^{\dagger}(\omega)\right]\vert0\rangle,\label{eq:state1}
\end{equation}
where $\psi_{a,f}(\omega)$, $\psi_{b,f}(\omega)$ are one-photon
wavefunctions. From Eqs.~(\ref{eq:P1}) and (\ref{eq:state1}), exploiting
Eqs.~(\ref{eq:bs}) and (\ref{eq:temporal}), we find 
\begin{equation}
P_{c,d}^{(1)}=\frac{1}{2}\pm\frac{1}{2}\int d\omega\left[\psi_{a,f}^{*}(\omega)\psi_{b,f}(\omega)+\text{c.c.}\right],\label{eq:P1cd}
\end{equation}
where we have imposed the normalization of the state, i.e. $\langle\psi_{f}\vert\psi_{f}\rangle=1$.
As an example let us consider the Quantum Sagnac experiment~\cite{bertocchi2006single}: a photon is prepared in a superposition of counter-propagating modes before interfering at the beam splitter (see supplementary material). Specifically, we consider the initial
state in Eq.~(\ref{eq:state1}) with the one-photon wavefunctions $\psi_{a,i}(\omega)=\psi_{b,i}(\omega)=g(\omega)$, where $g(\omega)$ is a Gaussian
with mean frequency $\mu$ and bandwidth $\sigma$. After the time-evolution
using Eq.~\eqref{eq:H2} we have the state in Eq.~(\ref{eq:state1})
with $\psi_{a,f}(\omega)=g(\omega)e^{i\omega\beta t_{f}}$ and $\psi_{b,f}(\omega)=g(\omega)e^{-i\omega\beta t_{f}}$.
Using Eq.~(\ref{eq:P1cd}), and making the further approximation
$\vert g(\omega)\vert^{2}\sim\delta(\mu-\omega)$, we find the single-photon detection
probability: 
\begin{alignat}{1}
P_{c,d}^{(1)} & =\frac{1}{2}\left(1\pm\text{cos}\left(\mu t_{s}\right)\right),\label{eq:P1e}
\end{alignat}
where $t_{s}={8\pi Af}/{c^{2}}$ is the classical Sagnac delay,
$f={\Omega}/{2\pi}$ is the rotation frequency, $A=\pi r^{2}$
is the encircled area, $r$ is the circle radius.

We next consider the two-photon state 
\begin{equation}
\vert\psi_{f}\rangle=\int d\omega_{1}\int d\omega_{2}\,\psi_{f}(\omega_{1},\omega_{2})\hat{a}^{\dagger}(\omega_{1})\hat{b}^{\dagger}(\omega_{2})\vert0\rangle,\label{eq:state2}
\end{equation}
where $\psi_{f}(\omega_{1},\omega_{2})$ is the two-photon spectrum.
From Eqs.~(\ref{eq:P2}) and (\ref{eq:state2}), exploiting Eqs.~(\ref{eq:bs}),
(\ref{eq:temporal}), we  find 
\begin{equation}
P^{(2)}=\frac{1}{2}-\frac{1}{2}\int d\omega\psi_{f}^{*}(\omega_{1},\omega_{2})\psi_{f}(\omega_{2},\omega_{1}),\label{eq:p2f}
\end{equation}
where we have imposed the normalization $\langle\psi_{f}\vert\psi_{f}\rangle=1$.
As an example we consider the Hong-Ou-Mandel experiment on a rotating
platform~\cite{restuccia2019photon}: two identical photons
counter-propagate before interfering at a beam-splitter (see supplementary material). The experimentalist
controls the initial time-delay $\delta t$ of the mode $\hat{a}$;
the initial state is given by Eq.~(\ref{eq:state2}) with the two-photon spectrum $\psi_{i}(\omega_{1},\omega_{2})=g(\omega_{1})g(\omega_{2})e^{-i\omega_{1}\delta t}$.
After the time-evolution we find the final state in Eq.~(\ref{eq:state2})
with 
\begin{equation}
\psi_{f}(\omega_{1},\omega_{2})=g(\omega_{1})g(\omega_{2})e^{-i\omega_{1}\delta t}e^{i\beta(\omega_{1}-\omega_{2})}.\label{eq:separable}
\end{equation}
Using Eq.~(\ref{eq:p2f}) we then immediately find the coincidence
probability: 
\begin{equation}
P^{(2)}=\frac{1}{2}-\frac{1}{2}e^{-\sigma^{2}(t_{s}-\delta t)^{2}}.\label{eq:p2_hom}
\end{equation}
where $t_{s}$ is the classical Sagnac delay.

In the previous paragraph we have
considered identical photons with a separable spectrum~\cite{hong1987measurement},
but one could also consider identical frequency-entangled photons.
For example, if we consider spontaneous parametric down conversion
(SPDC) type I two-photon generation~\cite{barbieri2017hong} we again
find the coincidence probability in Eq.~(\ref{eq:p2_hom}). It would
thus seem that entanglement in combination with rotational motion
leaves no trace on the photon coincidence rate, $P^{(2)}$, measurement. We now further explore
this question.

{\bf{Manifestation of entanglement through rotation}}.
Entanglement can manifest itself in a HOM coincidence rate measurement through
anti-coalescence, i.e. $P^{(2)}>0.5$. In particular, for
a completely anti-symmetric spectrum, i.e. $\psi(\omega_{1},\omega_{2})=-\psi(\omega_{2},\omega_{1})$,
one obtains perfect anti-coalescence, but even with a partially anti-symmetric
spectrum one can have $P^{(2)}>0.5$, thus witnessing entanglement.
As we show below, this manifestation of entanglement may be
susceptible to the motion of the interferometer. 
\begin{figure}
\includegraphics[width=4.5cm]{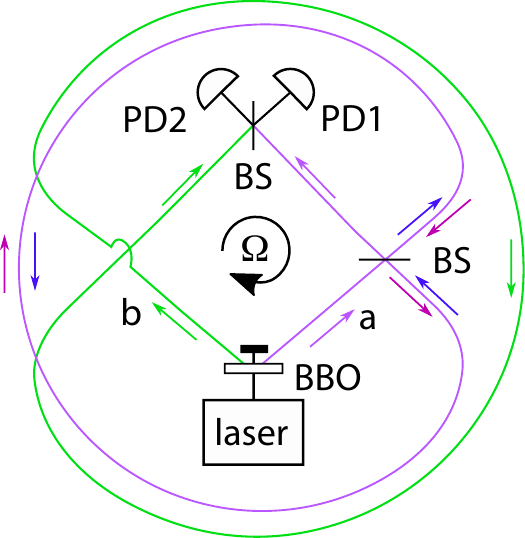}
\caption{Layout of proposed quantum Sagnac/Hong-Ou-Mandel interferometer on a rotating platform.
Two entangled photons are emitted from the BBO crystal: photon a (purple arrow) enters a Sagnac interferometer and exits towards the upper 50/50 beamsplitter (BS) where 2-photon HOM interference occurs with photon b (green arrow) that circles around the setup (in order to maintain the same overall path length as photon a). Coincidence counts are measured between detectors PD1 and PD2 as a function of the rotation frequency $\Omega$. }
\label{experiment} 
\end{figure}

We consider the experimental setup depicted in Fig.~\ref{experiment}.
As an initial state we consider a SPDC type I two-photon state (i.e.
two photons with the same polarisation) with spectrum 
\begin{equation}
\psi_{i}(\omega_{1},\omega_{2})=
\delta (\omega_{1}+\omega_{2}-2\mu)
g(\omega_{1})g(\omega_{2}),\label{eq:spdcIi}
\end{equation}
where we have omitted the normalization.
We note that the initial spectrum
in Eq.~(\ref{eq:spdcIi}) is completely symmetric, i.e. $\psi(\omega_{1},\omega_{2})=\psi(\omega_{2},\omega_{1})$,
and gives $P^{(2)}=0$ at $\Omega=0$. 
\begin{figure}[t]
\includegraphics[width=7cm]{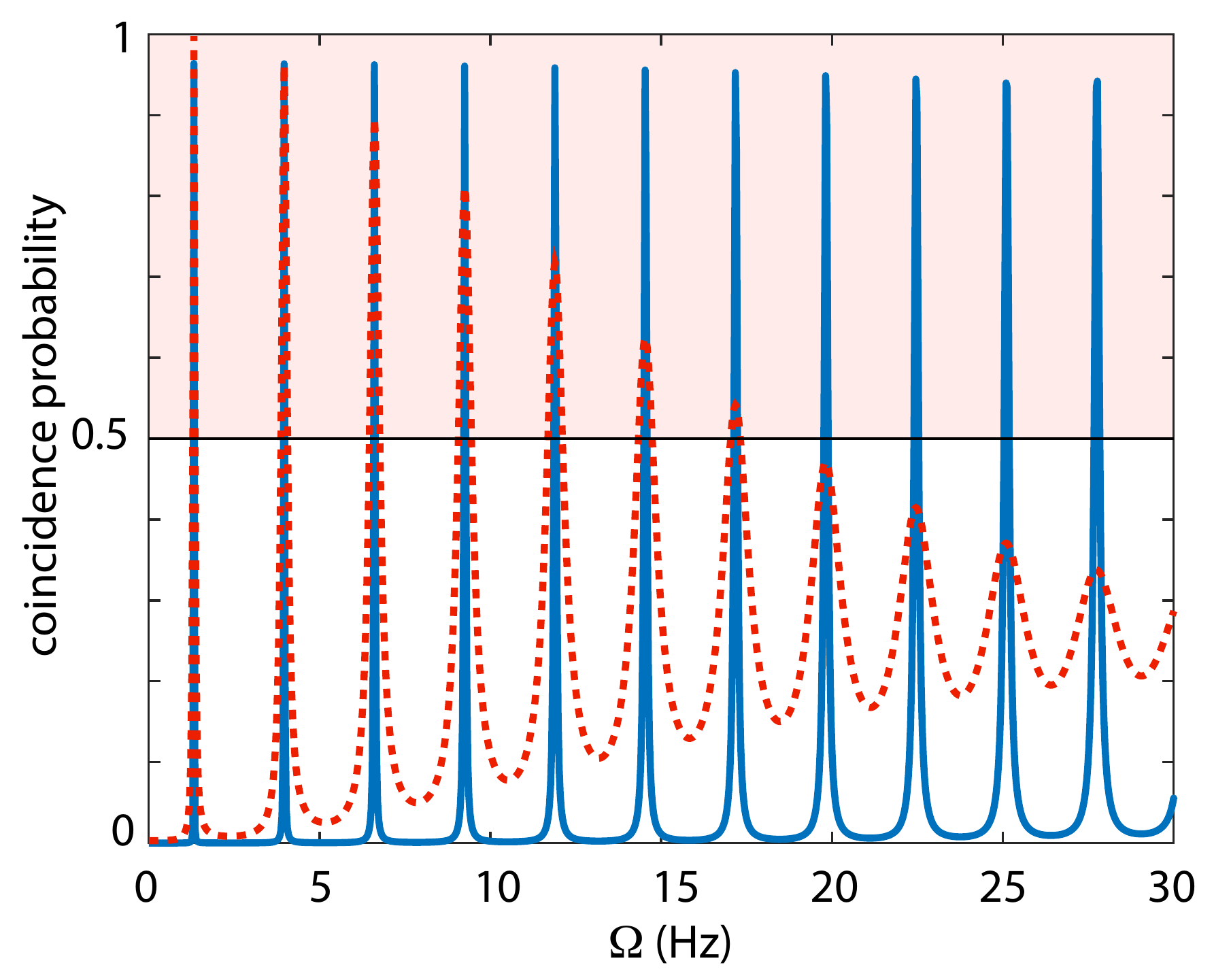}

\caption{Coincidence plot as a function of the angular frequency $\Omega=2\pi f$.
We have set the interferometer area $A=22.7\text{m}^{2}$, $\mu=2.36\times10^{15}\text{Hz}$, corresponding to a typical photon carrier wavelength of 800 nm.
Two curves are shown for two different bandwidths, $\sigma=1.47\times10^{13}\text{Hz}$ (blue solid curve) and $\sigma=1.18\times10^{14}\text{Hz}$ (dashed red curve), corresponding to 5 nm and 40 nm bandwidths, respectively. The shaded region corresponding to $P^{(2)}>0.5$ indicates the region where measurements imply photon entanglement. 
 }
\label{cp} 
\end{figure}

We now consider the same setup with the interferometer in a constant
rotational motion with frequency $\Omega\neq0$. The final spectrum
of the two-photon state changes to 
\begin{equation}
\psi_{f}(\omega_{1},\omega_{2})=\psi_{i}(\omega_{1},\omega_{2})\text{cos}(\beta\omega_{1}t_{f})e^{-i\beta\omega_{2}t_{f}},\label{eq:spdcIf}
\end{equation}
where the factor $\text{cos}(\beta\omega_{1}t_{f})$ results from
the interference of the mode $\hat{a}$; by `final state' we again
mean the state that arrives at the last beam-splitter. Using Eq.~(\ref{eq:p2f})
we then find the coincidence probability 
\begin{equation}
P^{(2)}=\frac{1}{2}-\frac{\cos(\mu t_{s})e^{-\frac{1}{8}\sigma^{2}t_{s}^{2}}+\frac{1}{2}\left(1+e^{-\frac{1}{2}\sigma^{2}t_{s}^{2}}\right)}{2(1+\cos(\mu t_{s})e^{-\frac{1}{8}\sigma^{2}t_{s}^{2}})}.\label{eq:pc}
\end{equation}
We have plotted $P^{(2)}$ as a function
of the angular frequency $\Omega$ in Fig.~\ref{cp} with the interferometer area $A=22.7\,\text{m}^{2}$ (assuming the photons travel through a 100 m long fibre, wound 35 times along a 0.9 m diameter loop) and $\mu=2.36\times10^{15}$ Hz, corresponding to a typical photon carrier wavelength of 800 nm.
Two curves are shown for two different photon bandwidths, $\sigma=1.47\times10^{13}$ Hz (blue solid curve) and $\sigma=1.18\times10^{14}$ Hz (dashed red curve), corresponding to 5 nm and 40 nm, respectively. The shaded region corresponding to $P^{(2)}>0.5$ indicates the presence of entanglement that manifests as photon \emph{anti-coalescence}. 
Short bandwidth, i.e. long coherence photons show a periodic series of revivals of entanglement with increasing rotation frequency. For larger photon bandwidths, i.e. shorter coherence lengths, increasing the relative photon delay by increasing the rotation speed leads to a reduction of the coincidence peak values and of the overall fringe visibility. This is a result of the loss of mutual coherence between the two interfering photons.

The anti-symmetrization of the photon spectrum, which leads to a modification
of the coincidence probability, is a direct consequence
of the non-inertial motion of platform. 
More generally, this shows that rotational
motion can activate dormant asymmetries in the experimental setup
leading to an anti-symmetric spectrum. It is also interesting
to consider an initial anti-symmetric spectrum $\psi_{i}$; the proposed
experiment shows that $\psi_{i}$ can become symmetrized during time-evolution,
fully concealing the anti-coalescence signature of entanglement.

These effects  can be traced to the impossibility of clock
synchronization along a closed loop on the rotating platform. In particular,
the Hamiltonian in Eq.~\eqref{eq:H2} can be linked to the effect
of clock desynchronization~\cite{gourgoulhon2016special}. It is
important to note that this is a genuine relativistic effect, which
is not expected to arise in a Newtonian theory, although it imprints
a non-negligible experimental trace in the regime typically associated
with the latter. This is different from the observer-dependent entanglement effect in non-inertial reference frames~\cite{fuentes2005alice}, expected to arise as a consequence of the Unruh radiation~\cite{davies1975scalar,unruh1976notes}  which vanishes at low accelerations. 

{\bf{Conclusions}}.
We have developed a formalism for describing interferometry experiments
on rotating platforms. We have first analyzed two recent photon-interferometry
experiments, namely, the quantum Sagnac and the Hong-Ou-Mandel experiment
on a rotating platform. We have then proposed a modified Hong-Ou-Mandel
interferometer where entanglement can be revealed or concealed depending
on the rotational frequency. These results indicate new directions
for investigating the notions of space and time as well as its consequences
in a quantum mechanical regime. 

{\bf{Acknowledgments}}.
The authors acknowledge support from the EU H2020FET project TEQ (Grant
No. 766900), from the EPSRC (UK, Grant No. EP/M009122/1) and from the European Union's Horizon 2020 research and innovation programme under grant agreement No 820392.


\begin{thebibliography}{10}

\bibitem{minkowski1909raum}
Hermann Minkowski.
\newblock Raum und zeit.
\newblock {\em Jahresbericht der Deutschen Mathematiker-Vereinigung, vol. 18,
  p. 75-88}, 18:75--88, 1909.

\bibitem{petkov2010minkowski}
Vesselin Petkov et~al.
\newblock {\em Minkowski spacetime: A hundred years later}, volume~11.
\newblock Springer, 2010.

\bibitem{von1910einige}
WA~Von~Ignatowsky.
\newblock Einige allgemeine bemerkungen zum relativit{\"a}tsprinzip.
\newblock {\em Verh. Deutsch. Phys. Ges}, 12:788--796, 1910.

\bibitem{liberati2013tests}
Stefano Liberati.
\newblock Tests of lorentz invariance: a 2013 update.
\newblock {\em Classical and Quantum Gravity}, 30(13):133001, 2013.

\bibitem{robertson1949postulate}
Howard~P Robertson.
\newblock Postulate versus observation in the special theory of relativity.
\newblock {\em Reviews of modern Physics}, 21(3):378, 1949.

\bibitem{michelson1887relative}
Albert~A Michelson and Edward~W Morley.
\newblock On the relative motion of the earth and of the luminiferous ether.
\newblock {\em Sidereal Messenger, vol. 6, pp. 306-310}, 6:306--310, 1887.

\bibitem{hong1987measurement}
Chong-Ki Hong, Zhe-Yu Ou, and Leonard Mandel.
\newblock Measurement of subpicosecond time intervals between two photons by
  interference.
\newblock {\em Physical review letters}, 59(18):2044, 1987.

\bibitem{mandel1999quantum}
Leonard Mandel.
\newblock Quantum effects in one-photon and two-photon interference.
\newblock In {\em More Things in Heaven and Earth}, pages 460--473. Springer,
  1999.

\bibitem{lyons2018attosecond}
Ashley Lyons, George~C Knee, Eliot Bolduc, Thomas Roger, Jonathan Leach, Erik~M
  Gauger, and Daniele Faccio.
\newblock Attosecond-resolution hong-ou-mandel interferometry.
\newblock {\em Science advances}, 4(5):eaap9416, 2018.

\bibitem{gourgoulhon2016special}
{\'E}ric Gourgoulhon.
\newblock {\em Special relativity in general frames}.
\newblock Springer, 2016.

\bibitem{sagnac1913ether}
Georges Sagnac.
\newblock L'{\'e}ther lumineux d{\'e}montr{\'e} par l'effet du vent relatif
  d'{\'e}ther dans un interf{\'e}rom{\`e}tre en rotation uniforme.
\newblock {\em CR Acad. Sci.}, 157:708--710, 1913.

\bibitem{sagnac1913preuve}
Georges Sagnac.
\newblock Sur la preuve de la r{\'e}alit{\'e} de l'{\'e}ther lumineux par
  l'exp{\'e}rience de l'interf{\'e}rographe tournant.
\newblock {\em CR Acad. Sci.}, 157:1410--1413, 1913.

\bibitem{post1967sagnac}
Evert~Jan Post.
\newblock Sagnac effect.
\newblock {\em Reviews of Modern Physics}, 39(2):475, 1967.

\bibitem{fink2017experimental}
Matthias Fink, Ana Rodriguez-Aramendia, Johannes Handsteiner, Abdul Ziarkash,
  Fabian Steinlechner, Thomas Scheidl, Ivette Fuentes, Jacques Pienaar,
  Timothy~C Ralph, and Rupert Ursin.
\newblock Experimental test of photonic entanglement in accelerated reference
  frames.
\newblock {\em Nature communications}, 8:15304, 2017.

\bibitem{fink2019entanglement}
Matthias Fink, Fabian Steinlechner, Johannes Handsteiner, Jonathan~P Dowling,
  Thomas Scheidl, and Rupert Ursin.
\newblock Entanglement-enhanced optical gyroscope.
\newblock {\em New Journal of Physics}, 21(5):053010, 2019.

\bibitem{restuccia2019photon}
Sara Restuccia, Marko Toro\v{s}, Graham~M. Gibson, Hendrik Ulbricht, Daniele
  Faccio, and Miles~J. Padgett.
\newblock Photon bunching in a rotating reference frame.
\newblock {\em Phys. Rev. Lett.}, 123:110401, Sep 2019.

\bibitem{bertocchi2006single}
Guillaume Bertocchi, Olivier Alibart, Daniel~Barry Ostrowsky, S{\'e}bastien
  Tanzilli, and Pascal Baldi.
\newblock Single-photon sagnac interferometer.
\newblock {\em Journal of Physics B: Atomic, Molecular and Optical Physics},
  39(5):1011, 2006.

\bibitem{tartaglia2015sagnac}
Angelo Tartaglia and Matteo~Luca Ruggiero.
\newblock The sagnac effect and pure geometry.
\newblock {\em American Journal of Physics}, 83(5):427--432, 2015.

\bibitem{da2001lectures}
Ana~Cannas Da~Silva and A~Cannas Da~Salva.
\newblock {\em Lectures on symplectic geometry}, volume 3575.
\newblock Springer, 2001.

\bibitem{woit2017quantum}
Peter Woit, Woit, and Bartolini.
\newblock {\em Quantum Theory, Groups and Representations}.
\newblock Springer, 2017.

\bibitem{padgett2008diffraction}
Miles~J Padgett.
\newblock On diffraction within a dielectric medium as an example of the
  minkowski formulation of optical momentum.
\newblock {\em Optics express}, 16(25):20864--20868, 2008.

\bibitem{barnett2010resolution}
Stephen~M Barnett.
\newblock Resolution of the abraham-minkowski dilemma.
\newblock {\em Physical review letters}, 104(7):070401, 2010.

\bibitem{glauber1963quantum}
Roy~J Glauber.
\newblock The quantum theory of optical coherence.
\newblock {\em Physical Review}, 130(6):2529, 1963.

\bibitem{legero2006characterization}
Thomas Legero, Tatjana Wilk, Axel Kuhn, and Gerhard Rempe.
\newblock Characterization of single photons using two-photon interference.
\newblock {\em Advances In Atomic, Molecular, and Optical Physics},
  53:253--289, 2006.

\bibitem{specht2009phase}
Holger~P Specht, J{\"o}rg Bochmann, Martin M{\"u}cke, Bernhard Weber, Eden
  Figueroa, David~L Moehring, and Gerhard Rempe.
\newblock Phase shaping of single-photon wave packets.
\newblock {\em Nature Photonics}, 3(8):469, 2009.

\bibitem{wang2006quantum}
Kaige Wang.
\newblock Quantum theory of two-photon wavepacket interference in a
  beamsplitter.
\newblock {\em Journal of Physics B: Atomic, Molecular and Optical Physics},
  39(18):R293, 2006.

\bibitem{blow1990continuum}
KJ~Blow, Rodney Loudon, Simon~JD Phoenix, and TJ~Shepherd.
\newblock Continuum fields in quantum optics.
\newblock {\em Physical Review A}, 42(7):4102, 1990.

\bibitem{fedrizzi2009anti}
Alessandro Fedrizzi, Thomas Herbst, Markus Aspelmeyer, Marco Barbieri, Thomas
  Jennewein, and Anton Zeilinger.
\newblock Anti-symmetrization reveals hidden entanglement.
\newblock {\em New Journal of Physics}, 11(10):103052, 2009.

\bibitem{barbieri2017hong}
Marco Barbieri, Emanuele Roccia, Luca Mancino, Marco Sbroscia, Ilaria Gianani,
  and Fabio Sciarrino.
\newblock What hong-ou-mandel interference says on two-photon frequency
  entanglement.
\newblock {\em Scientific reports}, 7(1):7247, 2017.

\bibitem{fuentes2005alice}
Ivette Fuentes-Schuller and Robert~B Mann.
\newblock Alice falls into a black hole: entanglement in noninertial frames.
\newblock {\em Physical review letters}, 95(12):120404, 2005.

\bibitem{davies1975scalar}
Paul~CW Davies.
\newblock Scalar production in schwarzschild and rindler metrics.
\newblock {\em Journal of Physics A: Mathematical and General}, 8(4):609, 1975.

\bibitem{unruh1976notes}
William~G Unruh.
\newblock Notes on black-hole evaporation.
\newblock {\em Physical Review D}, 14(4):870, 1976.

\end{thebibliography}


\begin{thebibliography}{1}

\bibitem{post1967sagnac}
Evert~Jan Post.
\newblock Sagnac effect.
\newblock {\em Reviews of Modern Physics}, 39(2):475, 1967.

\bibitem{gourgoulhon2016special}
{\'E}ric Gourgoulhon.
\newblock {\em Special relativity in general frames}.
\newblock Springer, 2016.

\bibitem{woit2017quantum}
Peter Woit, Woit, and Bartolini.
\newblock {\em Quantum Theory, Groups and Representations}.
\newblock Springer, 2017.

\bibitem{da2001lectures}
Ana~Cannas Da~Silva and A~Cannas Da~Salva.
\newblock {\em Lectures on symplectic geometry}, volume 3575.
\newblock Springer, 2001.

\bibitem{poisson2004relativist}
Eric Poisson.
\newblock {\em A relativist's toolkit: the mathematics of black-hole
  mechanics}.
\newblock Cambridge university press, 2004.

\bibitem{rizzi2002space}
Guido Rizzi and Matteo~Luca Ruggiero.
\newblock Space geometry of rotating platforms: an operational approach.
\newblock {\em Foundations of Physics}, 32(10):1525--1556, 2002.

\bibitem{tartaglia2015sagnac}
Angelo Tartaglia and Matteo~Luca Ruggiero.
\newblock The sagnac effect and pure geometry.
\newblock {\em American Journal of Physics}, 83(5):427--432, 2015.

\bibitem{bertocchi2006single}
Guillaume Bertocchi, Olivier Alibart, Daniel~Barry Ostrowsky, S{\'e}bastien
  Tanzilli, and Pascal Baldi.
\newblock Single-photon sagnac interferometer.
\newblock {\em Journal of Physics B: Atomic, Molecular and Optical Physics},
  39(5):1011, 2006.

\bibitem{restuccia2019photon}
Sara Restuccia, Marko Toro\v{s}, Graham~M. Gibson, Hendrik Ulbricht, Daniele
  Faccio, and Miles~J. Padgett.
\newblock Photon bunching in a rotating reference frame.
\newblock {\em Phys. Rev. Lett.}, 123:110401, Sep 2019.

\end{thebibliography}

\putbib
\end{bibunit}

\newpage
\begin{bibunit}
\begin{center}
\textbf{\large Supplementary material to ``Revealing and concealing entanglement with non-inertial motion''}
\end{center}

\setcounter{equation}{0}
\setcounter{figure}{0}
\setcounter{table}{0}
\setcounter{section}{0}
\setcounter{page}{1}
\makeatletter
\renewcommand{\thesection}{S\arabic{section}}
\renewcommand{\theequation}{S\arabic{equation}}
\renewcommand{\thefigure}{S\arabic{figure}}
\renewcommand{\bibnumfmt}[1]{[S#1]}
\renewcommand{\citenumfont}[1]{S#1}

\section*{S1: Laboratory reference frame}

In this section we briefly discuss the relation between the co-rotating reference frame and the inertial laboratory reference frame, focusing on the classical effects. We start by noting that the speed of light in the co-rotating reference
frame, where the co-rotating medium is stationary, is the same in
the co-rotating and counter-rotating directions; specifically we have
that the velocities are $\pm\frac{c}{n}$. Using the relativistic
velocity addition formula we find the corresponding velocities $u^{(\pm)}$
in the inertial laboratory frame:

\begin{equation}
u^{(\pm)}=\frac{\pm\frac{c}{n}+v}{1\pm\frac{c}{n}\frac{v}{c^{2}}}\approx\pm\frac{c}{n}+\alpha v,\label{eq:light_drag}
\end{equation}
where $v=r\Omega$, and $\alpha=1-\frac{1}{n^{2}}$. The corresponding
speeds are given by $v^{(\pm)}\approx\frac{c}{n}\pm\alpha v$. See
Fig.~1(b) for a graphical illustration of this light-dragging effect,
which coincides with the Fizeau effect, but could in principle differ~\cite{post1967sagnac}.

We assume an initial spatial-distance $L$ between the systems
and the detectors. We can convert $L$ into
a time-distance, i.e. the \emph{initial} time-distance (see Fig.~1(b1)), which is given by $t_{i}^{(\pm)}=\frac{L}{v^{(\pm)}}$. Furthermore,
exploiting Eq.~(\ref{eq:light_drag}), we find:

\begin{equation}
t_{i}^{(\pm)}\approx\frac{L}{c}n\mp\frac{vL}{c^{2}}\alpha n^{2}.\label{eq:tpmi1}
\end{equation}
However this is different from the \emph{actual} 
time it takes
the signals to reach the detectors (see Fig.~1 (b2)):
we need to take into account also the motion of the detectors. In
particular we have the condition $v^{(\pm)}t_{a}^{(\pm)}=L\pm vt_{a}^{(\pm)},$
which after some algebra readily gives

\begin{equation}
t_{a}^{(\pm)}=\frac{L}{v^{(\pm)}\mp v}\approx\frac{L}{c}n\pm\frac{Lv}{c^{2}}.\label{eq:tpma1}
\end{equation}
In this way we immediately recover the classic Sagnac delay given
by $t_{s}=t_{a}^{(+)}-t_{a}^{(-)}=\frac{2Lv}{c^{2}}$. Specifically,
to find the usual expression of the Sagnac delay we set $L=2\pi rN$,
where $N$ denotes the winding number, and define the encircled area
as $A=N\pi r^{2}$. Using $v=r\Omega=r2\pi f$ we then immediately
find~\cite{post1967sagnac,gourgoulhon2016special}: 
\begin{equation}
t_{s}=\frac{8\pi Af}{c^{2}}.
\end{equation}

From this discussion it is clear why the description in the co-rotating
reference frame is slightly more convenient: there only the non-inertial
motion of the detectors has to be taken into account (through the
Hamiltonians). On the other hand, in the laboratory inertial reference
frame, one has to account for the motion of the detectors as well
as of the medium (again through the Hamiltonians). In short, the advantage
of the co-rotating reference frame is the absence of the light-dragging
effect.

The Sagnac delay can be obtained also from the perspective of the
co-rotating observer where it arises from clock desynchronization~\cite{gourgoulhon2016special}. 

\section*{S2: Derivation of the Hamiltonian\label{sec:Derivation-of-Hamiltonian}}

In this section we derive the Hamiltonian for the experiments depicted
in Fig. 1(a): we consider a rotating platform, which spins at angular
frequency $\Omega$, and we restrict to the dynamics on a circle of
radius $r$, centered on the symmetry axis of the rotating platform.
Specifically, we will adopt the methods of representation theory~\cite{woit2017quantum}
and symplectic Hamiltonian mechanics~\cite{da2001lectures} to map
the time-evolution generator of the Poincaré algebra to a Hilbert
space operator. One could of course make an ad-hoc quantization in
a non-inertial reference frame, and obtain a Hamiltonian, but the
results might be inconsistent with basic symmetry requirements. Anyhow,
we choose the former method which constructs the Hamiltonian starting
from basic symmetry considerations of the Poincaré group. As argued
in the supplemental material S1 we will for convenience describe the
experiment in the co-rotating reference frame.

One typically starts describing the
experiments by setting up a chart, e.g. a Cartesian chart. We note
that the chosen spacetime coordinates critically reflect the motion
of the observer which affects the resulting description of the dynamics. For example,
two observers moving with different speeds or accelerations will use
different charts, and hence ascribe different energies to the same system, and hence care must be taken with the choice of the coordinate system. In the following we will assume that the detectors are stationary in the observer's chart: with this choice there is a simple relation between observables in the description and the quantities measured by the detectors.

We start by recalling the quantization procedure in an inertial reference
frame, i.e. with $\Omega=0$. In our case, we will use the polar chart for the laboratory inertial
reference frame. Specifically, the line-element in an inertial reference
frame expressed in the polar chart is given by

\begin{equation}
ds^{2}=c^{2}dt^{2}-r^{2}d\phi^{2},\label{eq:dsM}
\end{equation}
where $r$ is a constant in our case (we have restricted the analysis
to a $1+1$ spacetime). From the line-element in Eq.~(\ref{eq:dsM})
it is then possible to immediately read the time-evolution Killing
vector $(\frac{1}{c}\partial_{t})^{\mu}$~\cite{poisson2004relativist}.
The time-evolution Killing vector (more precisely a vector field)
is the tangent vector to the flow lines of a system in free motion.
Using the co-moment map we can then map the time-evolution
Killing vector to the Hilbert space Hamiltonian~\cite{woit2017quantum,da2001lectures}:
\begin{equation}
\partial_{t}\rightarrow\hat{H},
\end{equation}
where $\hat{H}$ is the \emph{representation} of the time-evolution generator
on the considered Hilbert space. For example, for a single mode $\hat{a}$
we would have $\hat{H}=\hbar\omega\hat{a}^{\dagger}\hat{a}$, where
$\omega$ is the frequency of the oscillator.

To apply the quantization procedure in a non-inertial reference frame,
i.e. with $\Omega\neq0$, we have to make an additional step: we have
to relate the co-rotating reference frame to the laboratory inertial
reference frame. There reason is simple: we do not know how to directly
quantize in a non-inertial reference frame, but only in the inertial
reference frame. We denote the laboratory (polar) and the co-rotating
(Born) coordinates by the unprimed ($x^{\mu}=(ct,r\phi)$) and primed
($x^{\mu'}=(ct',r\phi')$) labels, respectively. In particular, we
have the following relation~\cite{rizzi2002space}:

\begin{alignat}{1}
dt & =dt',\label{eq:time1}\\
d\phi & =d\phi'+\Omega dt'.\label{eq:phi1}
\end{alignat}
It is straightforward to find the corresponding transformation matrix

\begin{alignat}{1}
\frac{\partial x^{\mu}}{\partial x^{\mu'}}=\left[\begin{array}{cc}
1 & 0\\
\frac{r\Omega}{c} & 1
\end{array}\right],\label{eq:jacobian}
\end{alignat}
and to express the Minkowski metric in the two coordinates systems,
i.e.

\begin{alignat}{1}
ds^{2}= & c^{2}dt^{2}-r^{2}d\phi^{2}\label{eq:cartesian}\\
= & c^{2}(1-\frac{\Omega^{2}r^{2}}{c^{2}})dt'^{2}-2\Omega r^{2}dt'd\phi'-r^{2}d\phi'^{2}.\label{eq:Born}
\end{alignat}

From Eqs.~(\ref{eq:cartesian}) and (\ref{eq:Born}) we can immediately
find the relevant Killing vectors $(\frac{1}{c}\partial_{t'})^{\mu'}=(1,0)^{\top}$,
$(\frac{1}{c}\partial_{t})^{\mu}=(1,0)^{\top}$, and $(\frac{1}{r}\partial_{\phi})^{\mu}=(0,1)^{\top}$.
Using Eq.~(\ref{eq:jacobian}) we can then express $(\frac{1}{c}\partial_{t'})^{\mu'}$
in the laboratory coordinates, i.e.

\begin{alignat}{1}
(\frac{1}{c}\partial_{t'})^{\mu}=\frac{\partial x^{\mu}}{\partial x^{\mu'}} & (\frac{1}{c}\partial_{t'})^{\mu'},
\end{alignat}
which gives $(\frac{1}{c}\partial_{t'})^{\mu}=(1,\frac{r\omega}{c})^{\top}$,
and thus~

\begin{equation}
\partial_{t'}=\partial_{t}+r\Omega\frac{1}{r}\partial_{\phi}\,.\label{eq:kv}
\end{equation}
We have now expressed the time-evolution Killing vector $\partial_{t'}$,
which generates the dynamics in the co-rotating reference frame, in
terms of the Killing vectors $\partial_{t}$ and $\frac{1}{r}\partial_{\phi}$,
which generate time-evolution and space-translation in the inertial
laboratory reference frame, respectively. We can now map the latter
Killing vectors to operators on a Hilbert space using the usual prescription

\begin{alignat}{1}
\partial_{t} & \rightarrow\hat{H},\label{eq:h}\\
\frac{1}{r}\partial_{\phi} & \rightarrow\hat{P}.\label{eq:p}
\end{alignat}
Exploiting Eqs.~(\ref{eq:kv})-(\ref{eq:p}), we can now finally
write the time-evolution operator

\begin{equation}
\hat{H}_{\text{Born}}=\hat{H}+r\Omega\hat{P},\label{eq:born}
\end{equation}
which we name Born Hamiltonian. Note that Eq.~(\ref{eq:born}) captures
the idea that the dynamics on a rotating platform can be fully explained
in terms of the non-inertial motion of the detector~\cite{tartaglia2015sagnac}:
the term $r\Omega\hat{P}$ describes the non-inertial motion of the
detector, i.e. at each instant of time the detector is translated
with respect to the system which evolves freely on a circle.

We note that the transformation in Eq.~(\ref{eq:time1}) leaves the
time coordinate unchanged. This can be seen as a Galilean-type transformation
on a circle, which we now generalize to a Lorentz-type transformation.
In particular, we consider

\begin{alignat}{1}
dt & =\Gamma dt'+A\Gamma\frac{r^{2}\Omega}{c^{2}}d\phi',\\
d\phi & =B\Gamma d\phi'+\Gamma\Omega dt',
\end{alignat}
where $\Gamma=(1-(\frac{\Omega r}{c})^{2})^{-\frac{1}{2}}$. If we
set $A=1$ and $B=1$ the transformation is formally equivalent to
a Lorentz boost with speed $v=r\Omega$, while if we set $A=0$ and
$B=\Gamma^{-1}$ we obtain the transformation considered by Post~\cite{post1967sagnac}.
We find that the Hamiltonian is insensitive to the value of $A$,
but depends on the chosen value of $B$. In the following we set $B=1$
which leads to the Hamiltonian

\begin{equation}
\hat{H}_{\text{Born}}^{\text{rel}}=\Gamma(\hat{H}+r\Omega\hat{P}).\label{eq:bornRelS}
\end{equation}
Eq.~(\ref{eq:bornRelS}) can be seen as a relativistic Born Hamiltonian, which generalizes Eq.~(\ref{eq:born}).

We can also analyze non-uniform rotations using the above formalism
by considering a time-dependent angular frequency $\Omega_{t}$ (we remark that the time-evolution vector does not need to be generally
a Killing vector).  We
repeat the derivation in this section with the formal replacements
\begin{alignat}{1}
\Omega & \rightarrow\Omega_{t},\\
\Gamma & \rightarrow\Gamma_{t}=(1-(\frac{\Omega_{t}r}{c})^{2})^{-\frac{1}{2}}.
\end{alignat}
At the end we obtain in place of Eq.~(\ref{eq:bornRelS}) the following
Hamiltonian:

\begin{equation}
\hat{H}_{\text{Born}}^{\text{rel}}(t)=\Gamma_{t}(\hat{H}+r\Omega_{t}\hat{P}).
\end{equation}
In this case one expects two physical effects, one related to the \emph{(geometrical)} Sagnac phase, and a possible new contribution related to a \emph{dynamical phase}, which typically emerges in situations where there is a time-dependence in the Hamiltonian.  

\section*{S3: Notes on experimental setups\label{sec:Notes-on-experimental-setups}}

\begin{figure}[t!]
\vspace{0.5cm}
\begin{minipage}[t]{0.4\columnwidth}%
\includegraphics[width=0.8\columnwidth]{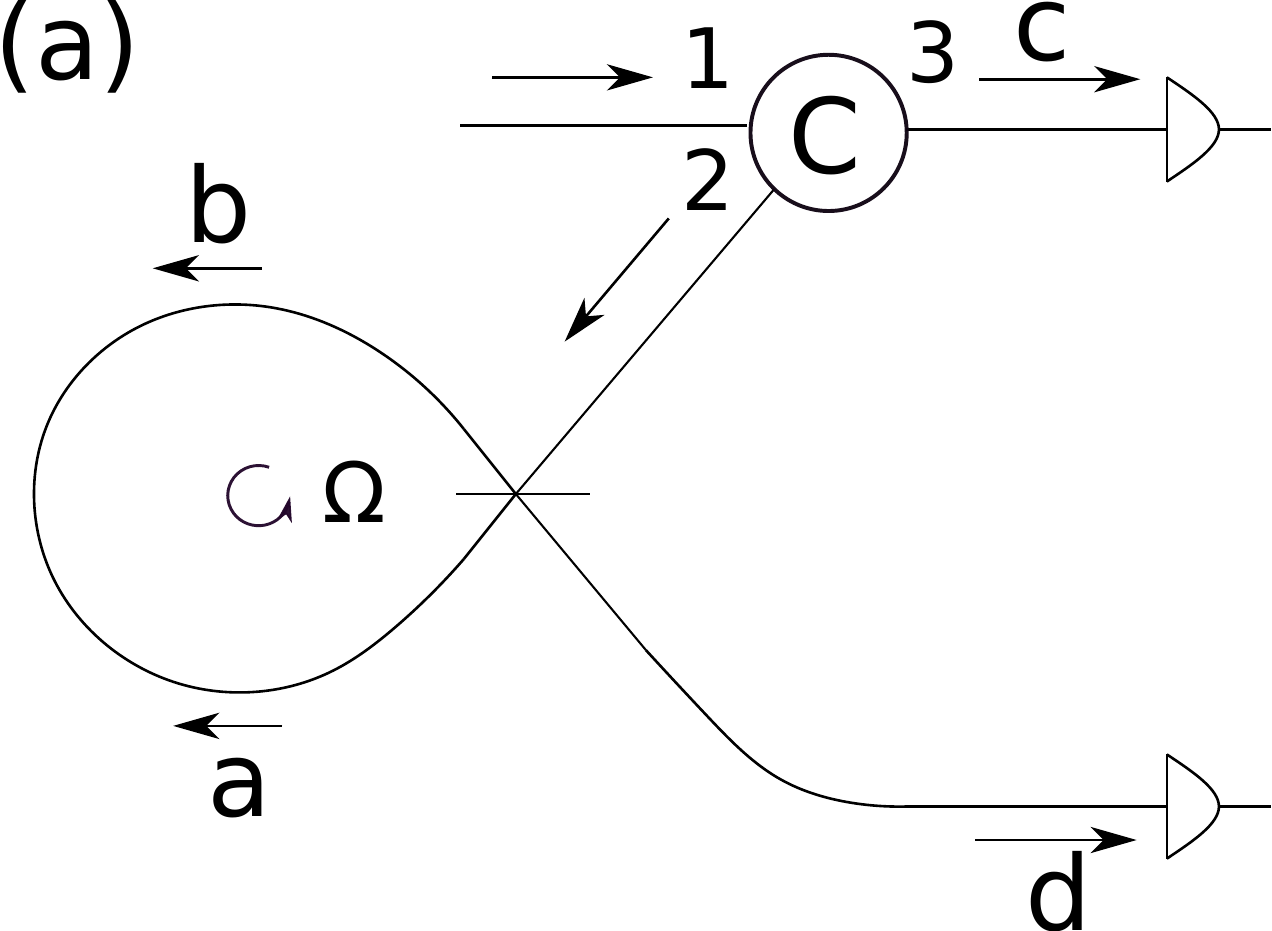}%
\end{minipage}\hspace{0.1cm}%
\begin{minipage}[t]{0.4\columnwidth}%
\includegraphics[width=1\columnwidth]{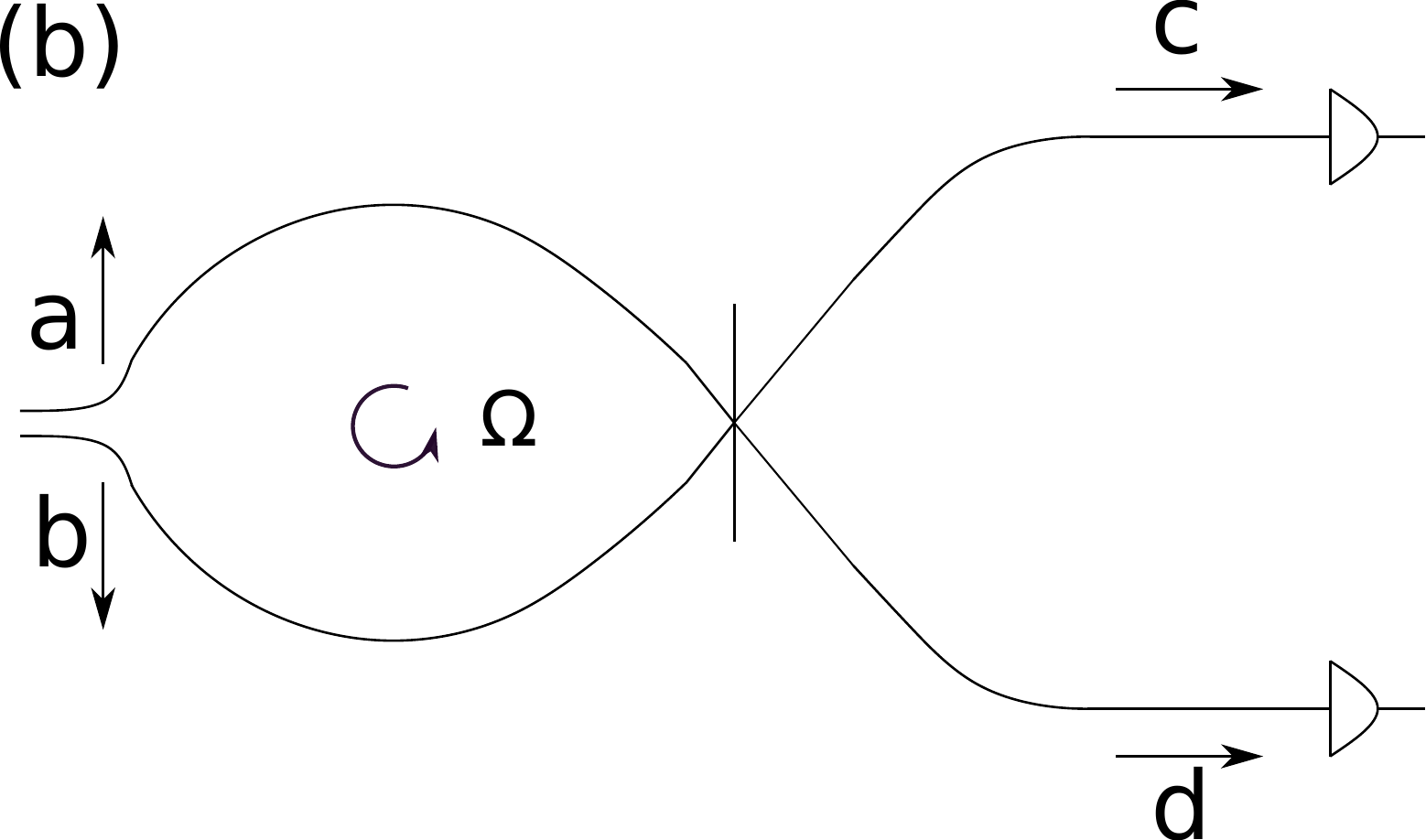}%
\end{minipage}\caption{(a) Quantum Sagnac experiment~\cite{bertocchi2006single}. The element
$C$ denotes the circulator which allows only the paths $1$ to $2$
and $2$ to $3$. (b) Hong-Ou-Mandel experiment on a rotating platform~\cite{restuccia2019photon}.}
\label{fig_schemes} 
\end{figure}

In this section we summarize the two experimental schemes that form the building blocks, conceptually as well as experimentally, for the new proposal to demonstrate how entanglement can be revealed or concealed with non-inertial motion. The Quantum Sagnac experiment is depicted in Fig.~\ref{fig_schemes}(a): a photon enters through the path 1, is directed into path 2 and then interferes for the first
time with the beam splitter. After evolving in counter-progating directions, the photon  then interferes again at the beam splitter, after which it is detected. The Hong-Ou-Mandel experiment on a rotating platform is depicted in Fig.~\ref{fig_schemes}(b): two identical photons
counter-propagate before interfering at a beam-splitter. One detects the arrival of the photons and extracts the coincidence probability, $P^{(2)}$. In both cases, the setups are placed on a rotating platform.

The proposal shown in Fig.~2 can be seen as a combination of the setups shown in Fig.~\ref{fig_schemes}, which can be exploited to gain an intuitive understanding of the results. For example, the asymptotic value for the coincidence probability in Fig.~3 can be intuitively understood in terms of Quantum Sagnac and the HOM setups. From Eq.~(19) we find at high rotation frequency the value $P^{(2)}\sim1/4$; this is halfway between a full dip, $P^{(2)}\sim 0$, and the case without bunching or anti-bunching, $P^{(2)}\sim1/2$. The paths denoted by the purple and blue arrows in Fig.~2 can be related to the Quantum Sagnac: at high rotation frequency the counter-propagating mode (blue arrow)  no longer interferes with the co-rotating modes due to non-overlapping frequency spectra. Loosely speaking, the counter-propagating mode (blue arrow) can be heuristically associated to half of the initial mode $\hat{a}$, while the other half of the mode $\hat{a}$ co-rotates (purple arrow) and interferes with the other co-rotating mode $\hat{b}$ (green arrow). As the frequency spectrum of the two co-rotating modes is identical, i.e. no time-delay exists between them, they  bunch together at the output ports of the beam-splitter. Hence the counter-rotating (co-rotating) part of the mode $\hat{a}$ is associated with $P^{(2)}\sim1/2$ ($P^{(2)}\sim0$). The coincidence probability at high rotation frequency can be thus seen as the average behaviour, i.e. $P^{(2)}\sim (0+1/2)/2=1/4$.

\putbib
\end{bibunit}

\end{document}